\documentclass[11pt]{article}
\pdfoutput=1
\usepackage{graphicx,rotating,hyperref,slashed,amsmath,xcolor,amssymb,amsfonts,colortbl,cite}
\hypersetup{colorlinks,bookmarksopen,bookmarksnumbered,
linkcolor=blus,pdfstartview=FitH,urlcolor=rossos,citecolor=verde}
\allowdisplaybreaks		
\def\beq{\begin{equation}}
\def\eeq{\end{equation}}

\def\bea{\begin{eqnarray}}
\def\eea{\end{eqnarray}}
\def\eg{{\it e.g.~}}

\def\d{{\rm d}}
\def\dd{{\rm d}}

\def\d{{\rm d}}

\def\nn{\nonumber}

\def\fr{\frac}

\def\0{{\boldsymbol 0}}

\def\lsim{\mathrel{\rlap{\lower3pt\hbox{\hskip0pt$\sim$}}
   \raise1pt\hbox{$<$}}}         
\def\gsim{\mathrel{\rlap{\lower4pt\hbox{\hskip1pt$\sim$}}
   \raise1pt\hbox{$>$}}}         

 \newcommand{\sfootnote}[1]{}
\definecolor{bluc}{cmyk}{1,1,0,0.1}
\definecolor{rossoCP3}{cmyk}{0,.88,.77,.40}
\definecolor{rosso}{cmyk}{0,1,1,0.4}
\definecolor{rossos}{cmyk}{0,1,1,0.55}
\definecolor{rossoc}{cmyk}{0,1,1,0.2}
\definecolor{verdes}{cmyk}{0.92,0,0.59,0.4}

\hypersetup{colorlinks, bookmarksopen, bookmarksnumbered,
citecolor=verdes, linkcolor=bluc, pdfstartview=FitH, urlcolor=rossos}

\newcommand{\mio}[1]{}

\definecolor{Gray}{gray}{0.95}

\usepackage{multicol}
\usepackage{color}
\definecolor{rosso}{cmyk}{0,1,1,0.4}
\definecolor{rossos}{cmyk}{0,1,1,0.55}
\definecolor{rossoc}{cmyk}{0,1,1,0.2}
\definecolor{blu}{cmyk}{1,1,0,0.3}
\definecolor{blus}{cmyk}{1,1,0,0.6}
\definecolor{bluc}{cmyk}{1,1,0,0.1}
\definecolor{verde}{cmyk}{0.92,0,0.59,0.25}
\definecolor{verdec}{cmyk}{0.92,0,0.59,0.15}
\definecolor{verdes}{cmyk}{0.92,0,0.59,0.4}

\def\circa#1{\,\raise.3ex\hbox{$#1$\kern-.75em\lower1ex\hbox{$\sim$}}\,}

\newcommand{\be}{\begin{equation}}
\newcommand{\ee}{\end{equation}}
\newfam\rsfsfam
\def\mathscr#1{{\fam\rsfsfam\relax#1}}

\def\circa#1{\,\raise.3ex\hbox{$#1$\kern-.75em\lower1ex\hbox{$\sim$}}\,}
\makeatletter

\def\hhref#1{\href{http://arxiv.org/abs/#1}{arXiv:#1}} 

\newcommand{\doi}[1]{\href{http://dx.doi.org/#1}{[doi]}}

\def\hhref#1{\href{http://arxiv.org/abs/#1}{arXiv:#1}} 
 
\def\art{\@ifnextchar[{\eart}{\oart}}
\def\eart[#1]#2#3#4#5#6{{\rm #2}, {\em #3 \bf #4} {\rm (#6) #5} ({\em #1})}

\def\article{\@ifnextchar[{\earticle}{\oarticle}}
\def\oarticle#1#2#3#4#5#6{{\rm #1}, {\em ``#6''}, {\rm #2 #3 (#5) #4}}
\def\earticle[#1]#2#3#4#5#6#7{{\rm #2}, {\em ``#7''}, {\rm #3 #4 (#6) #5}  [\hhref{#1}]}
\def\hepart[#1]#2{{\rm #2, \em#1}}
\def\heparticle[#1]#2#3{#2, {\em ``#3''} [\hhref{#1}]}

\newcounter{alphaequation}[equation]
\def\thealphaequation{\theequation\hbox to
0.6em{\hfil\alph{alphaequation}\hfil}}
\def\eqnsystem#1{
\def\@eqnnum{{\rm (\thealphaequation)}}
\def\@@eqncr{\let\@tempa\relax \ifcase\@eqcnt \def\@tempa{& & &} \or
  \def\@tempa{& &}\or \def\@tempa{&}\fi\@tempa
  \if@eqnsw\@eqnnum\refstepcounter{alphaequation}\fi
\global\@eqnswtrue\global\@eqcnt=0\cr}
\refstepcounter{equation} \let\@currentlabel\theequation \def\@tempb{#1}
\ifx\@tempb\empty\else\label{#1}\fi
\refstepcounter{alphaequation}
\let\@currentlabel\thealphaequation
\global\@eqnswtrue\global\@eqcnt=0 \tabskip\@centering\let\\=\@eqncr
$$\halign to \displaywidth\bgroup \@eqnsel\hskip\@centering
$\displaystyle\tabskip\z@{##}$&\global\@eqcnt\@ne
\hskip2\arraycolsep\hfil${##}$\hfil& \global\@eqcnt\tw@\hskip2\arraycolsep
$\displaystyle\tabskip\z@{##}$\hfil
\tabskip\@centering&\llap{##}\tabskip\z@\cr}
\def\endeqnsystem{\@@eqncr\egroup$$\global\@ignoretrue} \makeatother

\definecolor{fiorentina}{rgb}{.5,0,.5}

\begin{document}

\vspace{1truecm}
 
\begin{center}
\boldmath

{\textbf{\LARGE A new mechanism to
enhance primordial tensor fluctuations  in    single field  inflation}}

\unboldmath

\unboldmath

\bigskip\bigskip

\vspace{0.1truecm}
\date\today

{ {\color{black} Maria Mylova$^{a}$, Ogan \"Ozsoy$^{a}$, Susha Parameswaran$^{b}$, Gianmassimo Tasinato$^{a}$, Ivonne Zavala$^{a}$ }}
 \\[8mm]
{\it $^a$  Department of Physics, Swansea University, Swansea, SA2 8PP, UK}\\[1mm]
{\it $^b$ Department of Mathematical Sciences, University of Liverpool, Liverpool, L69 7ZL, UK}\\[1mm]

\vspace{1cm}

\thispagestyle{empty}
{\large\bf\color{blus} Abstract}
\begin{quote}
We discuss a new mechanism to enhance the spectrum of primordial tensor  fluctuations in single field inflationary scenarios. The enhancement relies on a transitory non-attractor inflationary phase, which amplifies the would-be decaying tensor mode, and
 gives rise to a growth of  tensor fluctuations at superhorizon scales. We show that the enhancement produced during this phase can be neatly treated via a  tensor duality between an attractor and non-attractor phase, which we introduce. We illustrate the mechanism and duality in a kinetically driven scenario of inflation, 
with non-minimal couplings between the scalar and the metric.
\end{quote}
\thispagestyle{empty}
\end{center}
\setcounter{page}{1}
\setcounter{footnote}{0}

\section{Introduction}
\label{sec-intro}

In standard single field inflation, the linearised dynamics of the scalar curvature fluctuation ${\cal R} $ is described by a quadratic action \cite{Mukhanov:1990me}
\be\label{scal-eq}
S_{R}\,=\,\int d \eta\,d^3 x\,\frac{z_S^2}{2}\,\left[ {\cal R'}^2-(\vec \nabla {\cal R} )^2\right]
\ee
with:
\be \label{scal-eq-ws}
z_S\,\equiv\,a\,\frac{\dot \phi}{H},
\ee
where $\phi$ is the homogeneous scalar field profile, and prime and dot indicate respectively derivatives along conformal and physical time. The combination $z_S$ 
  is known as the scalar pump field. If $z_S$ is an increasing function of time -- as in a slow-roll  regime, where the Hubble parameter and $\dot \phi$ are approximately constant -- then inflation is in an {\em attractor phase},  ${\cal R}$ is conserved at superhorizon scales and its spectrum is almost scale invariant. However, if $z_S$ is rapidly decreasing for  a brief
 interval,  then the inflationary evolution is no longer  an attractor: the would-be decaying mode of the curvature perturbation becomes dominant, and the power spectrum of modes leaving the horizon during this   {\em non-attractor phase} can be enhanced by several orders of magnitude in a short time interval. This can occur for example in models where the scalar derivative   rapidly decreases   for a short period, as in inflection point ultra slow-roll and in constant roll  inflationary systems (see e.g.  \cite{Inoue:2001zt,Linde:2001ae,Kinney:2005vj,Martin:2012pe,Motohashi:2014ppa,Yi:2017mxs,Dimopoulos:2017ged,Pattison:2018bct})
     or in the Starobinsky model with a rapid change in the potential slope \cite{Starobinsky:1992ts}. 
    Interestingly, although the non-attractor phase of inflation lies well outside the slow-roll regime, a duality exists    \cite{Wands:1998yp} which allows one to have an analytical  description   of the statistical features of the enhanced spectrum of fluctuations.
        Recently these scenarios have received a renewed  interest, since an  amplification of scalar 
         perturbations  can lead to    the production of primordial black holes   from single field inflation (see e.g. \cite{Garcia-Bellido:2017fdg,Sasaki:2018dmp,Carr:2016drx} for general reviews
   and  \cite{Garcia-Bellido:2017mdw,Germani:2017bcs,Motohashi:2017kbs,Ballesteros:2017fsr,Ezquiaga:2017fvi,Cicoli:2018asa,Ozsoy:2018flq,Biagetti:2018pjj,Saito:2008em,Kannike:2017bxn} for a  specific models). 
              
 \smallskip
       
Can we have a similar enhancement of  primordial tensor modes during a phase of
non-attractor single-field inflation? This question is phenomenologically interesting:  while   current
and forthcoming constraints from CMB polarization can probe the amplitude
of the  primordial tensor spectrum at very large CMB scales (see e.g. the reviews \cite{Chongchitnan:2006pe,Kamionkowski:2015yta}), 
 interferometers can probe a  stochastic background  of gravitational waves at much smaller scales (see
 the textbooks \cite{Maggiore:1900zz,Maggiore:2018sht}). Hence inflationary scenarios that enhance the spectrum of primordial tensor modes at interferometer scales   make predictions that are easier to test with interferometers instead of CMB experiments. So far, two main approaches  have been proposed. The first usually 
  exploits instabilities for  additional source fields during inflation: primordial gravity waves can be enhanced by coupling
  fields driving inflation with additional scalars \cite{Cook:2011hg,Senatore:2011sp,Carney:2012pk,Biagetti:2013kwa,Biagetti:2014asa,Goolsby-Cole:2017hod}, U(1) gauge vectors \cite{Sorbo:2011rz, Anber:2012du, Barnaby:2010vf , Barnaby:2012xt,Ozsoy:2017blg}, non-Abelian vector fields \cite{Maleknejad:2011jw,Dimastrogiovanni:2012ew,Adshead:2013qp,Adshead:2013nka,Obata:2014loa,Maleknejad:2016qjz,Dimastrogiovanni:2016fuu,Agrawal:2017awz,Adshead:2017hnc,Caldwell:2017chz,Agrawal:2018mrg},  or Standard Model fields \cite{Espinosa:2018eve}. 
   The second approach implements   
   space-time symmetry breaking during inflation. Ways to do so are scenarios of (super)solid inflation -- see e.g. \cite{Endlich:2012pz,Bartolo:2015qvr,Ricciardone:2016lym,Ricciardone:2017kre,Domenech:2017kno,Ballesteros:2016gwc,Cannone:2015rra,Lin:2015cqa,Cannone:2014uqa,Akhshik:2014bla} -- or massive gravity/bigravity models,  \cite{Lin:2015nda,Biagetti:2017viz,Dimastrogiovanni:2018uqy,Fujita:2018ehq}.  
  See e.g.  \cite{Bartolo:2016ami} for a more extensive survey of various  models proposed so far, focussing on the detectability of inflationary tensor modes with LISA.
   The aim of this work is to present a new mechanism to enhance the spectrum of  primordial tensor fluctuations in single field inflation,   
    which can be used to enhance spin 2 fluctuations 
     at arbitrary   scales. It is based on the hypothesis that the single field  inflationary dynamics encounters a brief  non-attractor phase  during its evolution:
     then the would-be decaying tensor mode grows at super-horizon scales, and enhance the tensor power spectrum. 
      The advantage of working in single field inflation is that we do not have to deal with backreaction of additional fields that can interfere with the inflationary dynamics. 
       We   consider a set-up with non-minimal couplings between the inflaton field and the metric, in order to have an adjustable function of time in the quadratic action for tensor fluctuations, which we shall use 
   to enhance the primordial tensor spectrum. We proceed
 as follows.
 \begin{itemize}
 \item In Section \ref{sec-enha} we study the second order action for primordial tensor fluctuations
 in single field inflation. We identify  conditions for obtaining a large enhancement of  the spectrum of primordial gravity waves,
  by exploiting a   non-attractor phase for tensors that amplify the would-be decaying tensor mode. These conditions are analogous to the requirements discussed in  various works, starting with \cite{Seto:1999jc,Leach:2000yw,Leach:2001zf}, for enhancing scalar modes during non-attractor phases, and motivate our search  for   models of inflation with specific non-minimal couplings of tensors to the inflationary scalar field.
 \item In general, it is difficult to have analytic control of the dynamics of fluctuations in a non-attractor phase. In 
  section \ref{sec-dua} we identify a criterium, which we call {\it tensor duality}, that ensures identical behaviour,    up to an overall factor, for the dynamics 
  of perturbations in two different  regimes of inflationary evolution. This is the generalization to the tensor case of the duality 
   discussed by Wands \cite{Wands:1998yp} for the scalar sector.  We determine the tensor dual  of a phase of standard slow-roll inflation,  which corresponds to a period of  non-attractor  inflation, with a scale invariant spectrum of tensor fluctuations    amplified with respect to the standard case.
  \item Using tensor duality as a guide, in section \ref{sec-modb} we build and analyse in detail a representative  model of single field kinetically driven inflation, belonging to the G-inflation set-up of \cite{Kobayashi:2011nu}, which is able to amplify tensor modes. 
    Our system is analogous to the Starobinsky model \cite{Starobinsky:1992ts}, where  instead of having discontinuities in the potential, we  have  a discontinuity in the kinetic functions which causes a short   non-attractor phase.
    Using tensor duality,  we are then able to analytically investigate the dynamics
   of fluctuations during the non-attractor era, showing that the amplitude of the spectra of tensor (and scalar) fluctuations
   increases by several orders of magnitude with respect to a standard slow-roll regime. 
 \item We conclude in section \ref{sec-out} with a discussion of possible future directions to explore, and provide technical appendices for some of the results of the main text. These include appendix B, where we use conformal and disformal transformations to translate the non-attractor evolution studied in the main text to an `Einstein frame', where the action for the tensor fluctuations takes the standard form. 
  \end{itemize}

\section{Enhancing    tensor  fluctuations in  single field  inflation}\label{sec-enha}

Consider linearised spin-2 tensor fluctuations around a FRW cosmological background,  defined as
\be\label{met}
d s^2\,=\,-d t^2+a^2(t)\,\left(
\delta_{ij}+h_{ij}
 \right)
 \,d x^i d x^j
\ee
where
 $h_{ij}$ is the transverse traceless spin-2 tensor perturbation.  At leading order in a derivative expansion, 
 the quadratic action for tensor perturbations can be expressed as  (see e.g. \cite{Kobayashi:2011nu}. From now on, we set $M_{Pl}=1$)
\bea
S_{T}&=&\fr{1}{8}\,\int d t \,d^3 x\,a^3(t)\,\,\left[ {\cal G}_T(t)\, \left( \partial_{t }
h_{ij} \right)^2-\frac{{\cal F}_T(t)}{a^2(t)}\left( \vec \nabla  h_{ij}  \right)^2\right]
\,,\nonumber \\
 \label{actssp}
&=&\fr{1}{2}\,\int d y \,d^3 x\,z_T^2(y)\,\left[  \left( \partial_{y }
h_{ij} \right)^2-\left( \vec \nabla  h_{ij}  \right)^2\right]\,.
\eea
The first line of this formula contains two functions of time ${\cal G}_T$, 
${\cal F}_T$ that characterise the tensor kinetic terms and {their time evolution} depends on the system under consideration. In the  second line of the previous expression we re-defined the time variable as
 \be \label{redtc}
 d t\,=\,a\,\left(\frac{{\cal G}_T}{{\cal F}_T}\right)^{1/2}\,d y
\,,
 \ee
 in order to  express the action for tensor fluctuations as the one for a free field  in a time dependent background. We also  introduced  the convenient combination
 \be \label{comz}
 z_T^2\,=\,\frac{a^2}{4}\,\sqrt{ {\cal G}_T\,{\cal F}_T} \,,
 \ee
 which we can call the  {\it tensor pump field} in analogy with the nomenclature used
  for scalar fluctuations.
{Notice that, in standard single field inflation, $ {\cal G}_T\,=\,{\cal F}_T \,=\,1$ and  $z_T^2\,\propto \,a^2(y)$ where $y\equiv \,\int \,d t/a(t)$ is the conformal time. However, in the presence of  non-minimal kinetic mixings between scalar and tensors, these functions ($\mathcal{G}_T$ and $\mathcal{F}_T$) can have a non-trivial time dependent profile, as we shall discuss at more length in what follows.} 
 

The equations of motion  in Fourier space
corresponding to the quadratic action \eqref{actssp} read (a prime indicates derivatives along $y$):

\be \label{eqev1a}
h''_{ij} +2\,\frac{ z_T'}{z_T}\, h'_{ij}
+{k^2}\,h_{ij}\,=\,0.
\ee
We focus here on super-horizon modes at large scales, defined as $k^2\ll |z_T''/z_T|$. In this case, the last term in eq.~\eqref{eqev1a} can be neglected, and
the super-horizon solution of eq.~\eqref{eqev1a} is given by
\be \label{solsh1}
h_{ij}(y) \,= \, {\cal C}_1+ {\cal C}_2 \,\int^y \frac{d y'}{z_T^2(y')}\,,
\ee
with ${\cal C}_{1,\,2}$ integration constants.
  If $z_T$ is a rapidly increasing function of time, as in standard slow-roll inflation, then the contribution of the second term is negligible. This implies  that
 tensor modes are  conserved at super-horizon scales, and the constant ${\cal C}_1$ is
fixed by matching this solution at horizon crossing
with the one for 
 sub-Hubble modes. But if  $z'_T/z_T$ changes sign in eq.~\eqref{eqev1a}, and
  $z_T$ becomes a rapidly decreasing
function of time -- even for a short time interval -- then the second term in \eqref{solsh1} increases with $y$
and can  become  dominant, enhancing the amplitude of tensor modes
with respect to the constant term ${\cal C}_1$.  Since the would-be decaying mode is no longer suppressed by inverse powers of the scale factor,
the system enters a {\it non-attractor regime} for the tensor sector:
\be \label{consNA}
 {\text{$\frac{z'_T}{z_T}\,<\,0$~~~~ $\Rightarrow$~~~~ non-attractor phase}}\hskip1cm\Rightarrow\hskip1cm{\text{amplification of tensor modes.}}
\ee
 The condition \eqref{consNA} can be achieved if the functions ${\cal G}_T$, ${\cal F}_T$ in eq.~\eqref{comz} have a strong time dependence. 
This very same mechanism is well known for the case of scalar fluctuations \cite{Leach:2000yw} 
 and has been exploited in recent literature for  producing primordial 
black holes from single field inflation during 
 a non-attractor phase
 \cite{Leach:2001zf}.
 As far as we are aware, we are the first to propose this effect as a mechanism
 to  enhance the inflationary tensor spectrum, i.e by amplifying the would-be decaying tensor mode, proportional
 to ${\cal C}_2$ in eq.~\eqref{solsh1}. 
 
 In general, the requirement  \eqref{consNA} for a non-attractor phase requires a violation of the slow-roll conditions, and the dynamics of 
 tensor fluctuations  
  is no longer analytically described by means of  the usual slow-roll formulae. However, based on an argument which we
  dub tensor duality, there exists a condition on the function $z_T$ of eq.~\eqref{comz} which allows us to have analytic control on the system during the non-attractor phase.  We discuss this duality in the next section.

\section{Duality for tensor  modes}
\label{sec-dua}

The idea of {\it duality} for scalar fluctuations has been well explored in the past, starting from \cite{Wands:1998yp}: see e.g. \cite{Finelli:2001sr,Gasperini:2002bn,Gratton:2003pe,Boyle:2004gv,Piao:2004uq,Allen:2004vz,Khoury:2008wj,Khoury:2010gw}. It allows one to identify scenarios that are able to produce 
a scale invariant spectrum of fluctuations without invoking a phase of quasi-de Sitter expansion, as in bouncing cosmologies (see e.g. \cite{Battefeld:2014uga,Brandenberger:2016vhg} for recent reviews on this broad topic). The same concept can be applied to the description of brief transient  phases of non-attractor evolution
during inflation, as first discussed in \cite{Leach:2000yw,Leach:2001zf}, explaining some of the key  features of scalar power spectra in these regimes. In this section we develop further this idea, extending it to the physics of tensor modes, with the aim of setting the stage for determining  scenarios that enhance tensor fluctuations during non-attractor inflationary regimes.

To investigate the concept   of tensor duality, let us focus on the quadratic action for tensor fluctuations as in 
eq.~\eqref{actssp}:
\be \label{actssp1}
S_{T}^{(2)}\,=\,\frac{1}{2}\int d y \,d^3 x\,z_T^2(y)\,\left[  \left( \partial_{y }
h_{ij} \right)^2-\left( \vec \nabla  h_{ij}  \right)^2\right].
\ee
We  canonically normalize the tensor field: 
\be
h_{ij} \,=\,
\frac{q_{ij}}{ z_T}
\,.
\ee
Then the quadratic action can be written in the standard form
\be \label{actssp3}
S_{T}^{(2)}\,=\,\frac12\,\int d y \,d^3 x
\,\left[  \left( 
q'_{ij} \right)^2- \,\left(\vec \nabla q_{ij}  \right)^2 +\frac{z_T''}{z_T}\,q_{ij}^2\right]\,,
\ee
corresponding to the action for 
  an harmonic oscillator with time dependent mass, which can be quantized and analytically investigated  under certain conditions. Any redefinition of 
  the function $z_T(y)$
which leaves the combination $z_T''/z_T$ invariant
{\it 
does not change the previous quadratic action}, and thus the equations of motion associated with 
 $q_{ij}$ as derived from eq.~\eqref{actssp3}  have
the same solutions. 
The most general such redefinition is the same as that discussed by Wands in the scalar sector \cite{Wands:1998yp}, and reads
\be \label{duomd}
\frac{\tilde z_T''}{\tilde z_T}\,=\, 
\frac{z_T''}{z_T}\hskip1cm {\text{iff}}\hskip1cm
\tilde z_T (y)\,=z_T(y)\left(\,{c}_1\,+{c}_2\,\,\int^{y}\,\frac{d y'}{z_T^2(y')}\right)
\,,\ee
for constants $c_{1,\,2}$. Since the action \eqref{actssp3} contains the same canonical variable after the redefinition of $z_T$, one can define a new tensor fluctuation $\tilde{h}_{ij}$ and relate it to the original perturbation $h_{ij}$ by rescaling $q_{ij}$ with the new function 
 $\tilde{z}_T$, namely:

\bea
\left\{ 
\begin{array}{l} 
h_{ij} = \,{q_{ij}}/{z_T}
\\ 
 \hskip3cm \Rightarrow\hskip1cm\tilde h_{ij}\,=\,({z_T}/{\tilde z_T})\,h_{ij}\,. \\ 
\tilde h_{ij} =\,{q_{ij}}/{\tilde z_T}
\end{array}
\right.  
\eea
We call the quantity $\tilde h_{ij}$ the {\it tensor dual} of $h_{ij}$.  The quadratic action describing the dynamics of 
 $\tilde h_{ij}$ has the same structure as eq.~\eqref{actssp1}, but contains  $\tilde z_T$ instead
 of $z_T$.
 %
  Since   both $h_{ij}$ and $\tilde h_{ij}$  are  associated with {the same canonical variable} $q_{ij}$, the quantity $\tilde h_{ij}$ has 
 the very same statistics as $h_{ij}$, and the corresponding power spectrum
is  related to  the original one by an overall rescaling:
\be \label{maph1}
{\cal P}_{\tilde h} \,=\,\left(\frac{z_T}{\tilde z_T}\right)^2\,{\cal P}_{ h}. 
\ee
This implies that if we have analytic control on the spectrum of perturbations $h_{ij}$ and their spectrum, we can 
easily control the spectrum for the dual fluctuations $\tilde h_{ij}$,  and if the ratio ${z_T}/{\tilde z_T}$ is 
large, the dual tensor spectrum is enhanced. This is what occurs for the tensor dual of spin-2 modes
in a slow-roll phase, as we  now  discuss. 

\subsection{The tensor dual of a slow-roll phase}\label{dualsr}

We   determine  the properties
of  the tensor  dual to a stochastic background of tensor modes produced during an inflationary  slow-roll    
%
 regime, where the background metric is quasi-de Sitter space and the scalar field profile is such that the functions ${\cal F}_T(t)$ and ${\cal G}_T(t)$ 
  are almost constant in time.
  The tensor spectrum in such slow-roll regime is  almost scale invariant: its amplitude at horizon exit is  
  given by
\bea\label{sts}
{\cal P}_{h} 
&=& \frac{2\,{\cal G}_T^{1/2}}{{\cal F}_T^{3/2}}\frac{H^2}{\pi^2}
\eea
where we neglect
the (weak) time dependence of the Hubble parameter and of the functions $\cal{G}_T$ and $\cal{F}_T$ (see e.g. \cite{Kobayashi:2011nu}
 for more complete expressions). 
 In this quasi-de Sitter slow-roll phase, the function $z_T$ as defined in eq.~\eqref{comz} is given by
 \be
 z_T^2\,=\,\frac{a^2}{4}\, \sqrt{ {\cal{G}}_T {\cal{F}}_T}\,\sim\,{\text{const.}}\,\times a^2
 \,.
 \ee 
On the other hand,
 the tensor  duality condition \eqref{duomd} between functions $\tilde z_T$ and $z_T$ 
 implies, approximating the background as pure de Sitter and taking ${\cal{G}}_T,\, {\cal{F}}_T$ constant, 
\bea
\partial_y \left( \frac{\tilde z_T}{z_T} \right)\propto \frac{1}{z_T^2}
\hskip0.5cm \Rightarrow \hskip0.5cm
\partial_t \left( \frac{\tilde z_T}{z_T} \right)\,\propto\,\frac{1}{a^3}
\hskip0.5cm \Rightarrow \hskip0.5cm
\frac{\tilde z_T}{z_T} \,\propto\,\frac{1}{a^3}
\hskip0.5cm \Rightarrow \hskip0.5cm
{\tilde z_T^2} \,\propto\,\frac{1}{a^4}
\label{duaco1}
\eea 
where we used relation \eqref{redtc} to connect the $y$ and $t$ time variables, in terms of quantities defined in the slow-roll phase. 
Hence we learn that while the slow-roll phase is an attractor, with $z_T$ increasing with time, in the dual tensor phase the function $\tilde z_T$
decreases: we are in a non-attractor regime in which  tensor fluctuations can grow at superhorizon scales,  as discussed
in Section \ref{sec-enha}. 
  On the other hand, in the dual tensor phase, we have $\tilde h_{ij}\,=\,(z_T/\tilde z_T) \,h_{ij}$. The spectrum of tensor fluctuations continues to be almost scale invariant,  with an enhanced amplitude given by
 \be \label{tensenha}
 {\cal P}_{\tilde h} \,=\,\left(\frac{z_T}{\tilde z_T}\right)^2\,{\cal P}_{ h} \,\propto\,a^6\, {\cal P}_{ h}.
 \ee
 Since the scale factor $a$ is an  increasing function of time, 
this mechanism allows one to considerably amplify the tensor spectrum
 in the regime where eq.~\eqref{duaco1} holds,
 maintaining it almost  flat. The mechanism is the analog  for
 the tensor spectrum of
 an idea   introduced by Wands \cite{Wands:1998yp} and well explored in the scalar sector \cite{Biagetti:2018pjj}. 
Given that during slow-roll $({\cal F}_T\,{\cal G}_T)^{1/2}\,\simeq\,{\text{const.}}$, the condition \eqref{duaco1} implies that in the tensor dual phase the functions  $\tilde
{\cal F}_T$, $\tilde {\cal G}_T$ satisfy the relation
\be
\sqrt{\tilde{{\cal F}}_T\, \tilde{{\cal G}}_T}\,\propto\,\frac{1}{a^6},
\label{con2o}
\ee
which defines a non-attractor regime for tensor fluctuations\footnote{\label{fot1} As mentioned in the text, tensor duality is a closed relative
of the scalar duality first introduced in \cite{Wands:1998yp} for standard single field inflation. Starting from an action for scalar fluctuations as in equation  \eqref{scal-eq}, the 
 scalar duality states that  the statistics  of  scalar fluctuations do not change in regimes related by a condition 
\be
\tilde z_S(\eta)\,\propto\,z_S(\eta) \, \int^\eta \frac{d \eta'}{z^2_S(\eta')}\,.
\hskip1cm 
\ee
In the dual of a slow-roll, quasi-de Sitter phase of expansion with $\dot \phi\,=\,$const.,
 the scalar velocity must decrease as
$
 \dot{\tilde \phi} \,\propto\,
 \dot \phi \,\int \frac{d t'}{a^3 (t')\,\dot \phi (t')} 
 \,\propto\,
 \frac{1}{a^3} \,,
 $
 precisely the behaviour one encounters in a non-attractor, ultra slow-roll regime of inflation. In the scalar  dual of a slow-roll phase, scalar fluctuations are enhanced by a factor ${\cal P}_{\tilde {\cal R}} \,\propto\,a^6\, {\cal P}_{{\cal R}}$.
}.
In the next section, we present an explicit example that is
able to realise such a regime during a short phase of non-attractor
inflationary evolution.

\section{Amplifying tensor modes and realising tensor duality in single field inflation  }\label{sec-modb}

\subsection{Our aims} \label{ref-aims}

We now seek a realization of the amplification mechanism and the tensor duality of Sections \ref{sec-enha} and  \ref{sec-dua}  in a single field inflationary system,  whose background evolution is 
controlled by the  scalar $\phi$.  When non-minimal derivative couplings between scalar and metric
 are present, the functions ${\cal F}_T$ are ${\cal G}_T$ can have non-trivial time dependent 
profiles: 
%
%
%
 we look for a situation
where they can be
  expressed
as
\be \label{satc1}
{\cal G}_T\,\propto\,\frac{\dot \phi^2}{H^2} \hskip0.5cm,  \hskip0.5cm
{\cal F}_T\,\propto\,\frac{\dot \phi^2}{H^2}
\hskip0.5cm \Rightarrow  \hskip0.5cm \sqrt{{\cal F}_T {\cal G}_T} \,\propto\,\frac{\dot \phi^2}{H^2}
\hskip0.5cm,  \hskip0.5cm
z_T\sim a \,\frac{\dot \phi}{H} \,.
\ee
The reason for this choice is to `mimic' the behavior of the action for scalar fluctuations, as briefly reviewed
in section \ref{sec-intro} and footnote \ref{fot1}. Indeed, if  condition \eqref{satc1} is satisfied, we can apply the same results designed to enhance  fluctuations in the scalar sector (see e.g. \cite{Garcia-Bellido:2017mdw,Germani:2017bcs,Motohashi:2017kbs,Ballesteros:2017fsr,Ezquiaga:2017fvi,Cicoli:2018asa,Ozsoy:2018flq,Biagetti:2018pjj} for recent studies) to the tensor sector. In particular, we are interested in scenarios in which a phase of  de Sitter expansion (where
 $H$ and $\dot \phi$ are approximately  constant) is briefly interrupted by a phase of non-attractor inflation with de Sitter expansion, but where
 $\dot \phi\sim 1/a^3$. In this case, one passes from $ \sqrt{{\cal F}_T {\cal G}_T} = \,{\text{const.}}$ during slow-roll to   $\sqrt{{\cal F}_T {\cal G}_T} \propto \,{1}/{a^6}$ during non-attractor inflation, precisely what we need to amplify the tensor modes and realise the tensor duality:
  see eqs.~\eqref{con2o}, \eqref{satc1}.

 \smallskip
 
 The simplest possibility for having a regime where $\dot \phi$ transiently decreases  during inflation is the
 scenario of Starobinsky \cite{Starobinsky:1992ts} (see Appendix A of \cite{Biagetti:2018pjj} for a detailed analysis of this scenario), in which a linear inflationary potential $V(\phi)$ is continuous but has an abrupt change
 of slope for a certain value of the scalar field. In this case, the scalar field velocity $\dot \phi$  rapidly changes during a short 
 fraction of the inflationary 
 period to adapt its value 
  from the first to the second slow-roll regimes characterised by different potentials.
   During the transition,  its value decreases as desired, $\dot \phi\sim 1/a^3$, for an appropriate
  choice of the parameters involved. Whilst the Starobinsky model does indeed lead to an enhancement of scalar fluctuations, as 
  ${\cal F}_T  \sim 1 \sim {\cal G}_T$ tensor modes remain small.
  
  Our purposes on this section are the following.
  \begin{enumerate}
\item   Design a system of kinetically driven inflation where  the scalar evolution undergoes a brief non-attractor phase during which $\dot \phi\sim 1/a^3$. We build
a version of Starobinsky model \cite{Starobinsky:1992ts} based on Horndeski Lagrangians and non-standard kinetic terms, where non minimal derivative couplings between metric and scalar field allow us to have a
 rich  dynamics for tensor fluctuations and appropriate time profiles for the functions ${\cal F}_T$, ${\cal G}_T$. 
 \item Select the parameters of the system  to ensure that condition \eqref{satc1} is satisfied, so that during the short non-attractor regime 
  in which $\dot \phi\sim 1/a^3$, tensor modes are enhanced and the tensor duality applies.
\item Ensure that no ghost or gradient instabilities occur both in the tensor and scalar sectors of fluctuations.
\item However, {\bf we do not aim} at building a realistic inflationary model which matches with CMB observations at large scales and has
a realistic exit from inflation. Our purpose is specifically to show that our
 mechanism for enhancing tensor fluctuations can be realised in a toy model for inflation, while a more realistic set-up will be explored elsewhere.  
\end{enumerate}
 
 \subsection{The model}
 
 We build a model of single field inflation in a general set-up with non-minimal derivative couplings
 between scalar and tensors, in order to have the opportunity to enhance tensor modes and realise the tensor duality.
 The framework we work with is  Horndeski theory, which corresponds to the most general covariant scalar-tensor
 system with second order equations of motion, with Lagrangian density:
%
%
%
%
%
%
%
\bea
{\cal L}_{\text{tot}}&=&{\cal L}_2+{\cal L}_3+{\cal L}_4+{\cal L}_5
\,,
\\
{\cal L}_2&=&G_2\,
\,,
\\
{\cal L}_3&=&-G_3\,\Box \phi\,
\,,
\\
{\cal L}_4&=&G_{4}\,R+G_{4 X} \left[ \left( \Box \phi\right)^2 -\left( \nabla_\mu \nabla_\nu \phi \right)^2\right]
\,,
\\
{\cal L}_5&=&
G_5\,G_{\mu\nu}\,\nabla^\mu\nabla^\nu\,\phi-\frac{G_{5X}}{6}
\,
\left[
\left( \Box \phi \right)^3-3 \Box \phi\,\left(\nabla_\mu \nabla_\nu \phi \right)^2
+2 \left(\nabla_\mu \nabla_\nu \phi \right)^3
\right]
\,.
\eea
The quantities $G_a\,=\,G_{a} (\phi,X)$ ($a=1,\dots 5$) are arbitrary functions of the scalar field $\phi$ and 
\be
 X\,=\,-\,\frac12 \,\partial_\mu \phi \partial^\mu \phi\,,
 \ee
   $R$ is the Ricci tensor, $G_{\mu\nu}$ is the Einstein tensor, 
   and $G_{a\,X} = \partial G_a/\partial X$.
   Scenarios of single field inflation with standard kinetic terms are described by the following choice of the functions $G_i$,
\be\label{SI}
{\text{Standard inflation:}} \hskip0.5cm G_2\,=\,X-V(\phi), \hskip0.5cm\hskip0.5cm G_4\,=\,\frac{1}{2},
 \hskip0.5cm\hskip0.5cm G_3\,=\,G_5\,=\,0\,,
\ee
with $V(\phi)$ the inflationary potential, and recall that we set $M_{Pl}=1$: in this case, the Lagrangian ${\cal L}_4$
corresponds to the standard Einstein-Hilbert action. Single field  inflationary systems based on Horndeski
and Galileon Lagrangians
have been studied in many works, starting from Galileon inflation \cite{Burrage:2010cu} and the more general G-inflation \cite{Kobayashi:2010cm,Kobayashi:2011nu} scenarios. Scenarios of ultra slow-roll, non-attractor G-inflation have been discussed in \cite{Hirano:2016gmv}, concentrating on the dynamics of  scalar fluctuations.  Other single
field  models
of kinetically driven non-attractor inflation have been explored in \cite{Chen:2013aj,Chen:2013eea} (see also \cite{Cai:2017bxr}), focussing especially on the enhancement of scalar non-Gaussianity in the squeezed limit.

In what follows, for simplicity we make the hypothesis that all functions $G_a$ depend on the kinetic functions $X$
only,
\be
G_a\,=\,G_a (X)
\ee
and we focus on 
scenarios of {\em kinetically driven inflation}, where the inflationary evolution is driven not by a potential, but by the non-linear structure of the 
kinetic sector.

 We build a version 
of the Starobinski model \cite{Starobinsky:1992ts} in this context, by choosing the following structure for the functions $G_a(X)$:
%
%
%
%
 \bea\label{GF}
\nn G_2^{(i)}&=&\rho_i X+\frac{\sqrt 2}{3}\,H_0^2\,\alpha_i\,\sqrt{X}
 -V_i  \,,
 \\
\nn G_3^{(i)}&=&\frac{\sqrt{2}}{3\,H_0}\,\delta_i\,\sqrt{X} \,,
 \\
\nn G_4^{(i)}&=&-\frac{\beta_i}{6 H_0^2}\,X   \,,
 \\
 G_5^{(i)}&=&\frac{\sigma_i}{\sqrt{2}\,H_0^3 }\,\sqrt{X}     \,,    
 \eea
 where $i=1,2$, denote the two different phases of inflation we discuss below,  and $H_0$ is a mass scale. These functions depend on a set of dimensionless parameters ${\alpha_i,\beta_i,\,\delta_i,\,\rho_i,\,\sigma_i,\,V_i}$, needed  to satisfy all the  conditions discussed at the end of the Section \ref{ref-aims}.
 We make the hypothesis that these parameters   {\it change their magnitude} at a given time $t=t_0$, 
 making the above functions discontinuous:
 \be
\left(
{\alpha_i,\beta_i,\,\delta_i,\,\rho_i,\,\sigma_i,\,V_i }
\right)
\,=\, \left\{ \begin{array}{l} 
{\alpha_1,\,\beta_1,\,\delta_1,\,\rho_1,\,\sigma_1,\,V_1\,,} \hskip3cm t\,<\,t_0\,,
\\ 
\\
 {\alpha_2,\,\beta_2,\,\delta_2,\,\rho_2,\,\sigma_2,\,V_2\,,} \hskip3cm t\,>\,t_0\,.
 \end{array} \right. 
 \ee
  Nevertheless, by selecting appropriately the integration constants, the physical metric
 and field velocity
  can be made continuous. The scalar field velocity on the other hand, will have to abruptly change its slope during the inflationary evolution, in order to  accommodate the parameter
 discontinuities\footnote{{E.g. a change in the parameter $\beta_i$ could be caused in a string theory set up by a change in the volume of the extra compactified dimensions.  }Notice that the scenario could be improved smoothing out the discontinuities by means of some steep functions,
  as done in  \cite{Biagetti:2018pjj} for the Starobinsky model.}: we use precisely   this effect to enhance tensor fluctuations in this set-up. It is important to point out that the choice of functions $G_a$ of equation \eqref{GF} is designed to study the amplification of tensor modes, but the phase of kinetic domination should end before inflation terminates, in order to have a realistic exit from inflation and ensure that a standard Einstein-Hilbert term (with a constant $G_4$) is obtained after inflation. This condition might be achieved enriching the system and allowing for explicit dependence on $\phi$ of the free functions involved: this interesting topic goes beyond the scope of this work, and we leave it for future investigations. 
   
\subsection{Background evolution}

For the choices of functions $G_a(X)$ in \eqref{GF}, the background equations\footnote{Some details on the background equations of the scalar-tensor system we are considering can be found in the Appendix A.} for the scalar and the metric read, respectively
 
\be
\frac{d}{d t} \left\{ a^3(t) \left[ \alpha_i\,\frac{H_0^2}{3}+ \left( -\rho_i+
\delta_i\,\frac{H(t)}{H_0}+
\beta_i \frac{H^2(t)}{H_0^2}
+\sigma_i \frac{H^3(t)}{H_0^3}
\right) \dot \phi\right]\right\}\,=\,0
\ee
and
\be \label{eomH}
V_i\,=\,\frac{\dot \phi^2(t)}{2} \left( -\rho_i+2\,\delta_i\,\frac{H(t)}{H_0}+3\beta_i \frac{H^2(t)}{H_0^2}
+4\sigma_i \frac{H^3(t)}{H_0^3}
\right) 
\ee
where $i=1,2$ correspond to the two phases of evolution, before and after the transition at $t=t_0$.   
We focus on solutions where the scalar field velocity is monotonic  (with convention $\dot{\phi} < 0$) and the scale factor is exponentially increasing (de Sitter space)
with constant Hubble parameter $H_0$ during the entire inflationary evolution.

 We choose integration constants and parameters so that 
 in the first part of the evolution, $t\le t_0$,  the scalar field velocity is constant. In the second part of the evolution ($t\ge t_0$) the scalar velocity can vary,
 and we impose continuity of the quantity  $\dot \phi$ at $t=t_0$.  
 The solution for the scalar equation with the desired property is 
\bea\label{dphi}
\nn \dot{\phi}&=&-\frac{H_0^2\,\alpha_1}{3 \left(-\rho_1+\delta_1+\beta_1+\sigma_1 \right)}\hskip1cm t\le t_0\,,
\\
\dot{\phi}&=&
 -\frac{H_0^2}{3 \left(-\rho_2 +
\delta_2 
+
\beta_2 +\sigma_2 \right)} \, 
\left[
\alpha_2\,\left(1-\frac{1}{a^3(t)} \right)+\frac{\alpha_1}{\, a^3(t)}\,\frac{(-\rho_2+ \delta_2 +\beta_2+\sigma_2)}{(-\rho_1+ \delta_1 +\beta_1+\sigma_1)}
\right]
%
%
\hskip1cm t\ge t_0
\,, \nonumber
\\
\eea
where $H(t)\,=\,H_0$ is the constant Hubble
parameter, and   the scale factor reads
\be
a(t)\,=\,e^{H_0 (t-t_0)}\,.
\ee

\begin{figure}[h!]
\begin{center}
\includegraphics[width = 0.7 \textwidth]{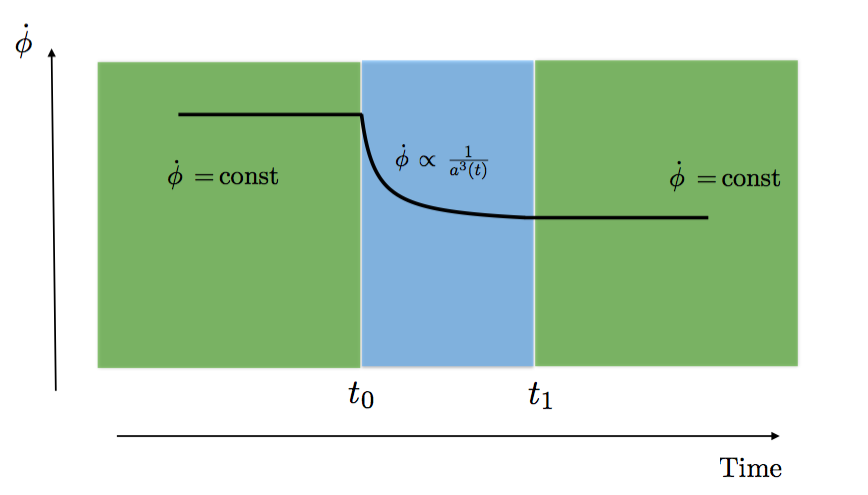}
 \caption{\it 
Schematic behaviour of the scalar field derivative in our system. Notice the intermediate non-attractor phase $\dot \phi\sim1/a^3$.  
During the entire inflationary  evolution the background geometry corresponds to pure de Sitter space. }
\label{fig:scal}
\end{center} 
\end{figure}

From now on we make
 the hypothesis that
 $\alpha_2$ is very small such that for $t>t_0$ the system enters  a short non-attractor phase, with rapidly decreasing field velocity
 \be
 \dot \phi \propto \frac{1}{a^3}\,,
 \ee
  lasting 
  from $t=t_0$ until $t\,=\,  t_1$ when 
\be \label{defttil}
 \alpha_2\,\approx
\frac{\alpha_1}{\, a^3(t_1)}\,
\frac{(-\rho_2+ \delta_2 +\beta_2+\sigma_2)}{(-\rho_1+ \delta_1 +\beta_1+\sigma_1)}
%
%
%
\,.
\ee
%
For $t>   t_1 $ the scalar time derivative returns to being a constant, with value
\bea \label{scal3p}
\dot{\phi}&=&-  \frac{H_0^2\,\alpha_2}{ 3 \left(-\rho_2 +
\delta_2
+
\beta_2+\sigma_2 \right)} \, .
\eea
The last phase of slow-roll evolution, $t>t_1$, is the least interesting one for our purposes -- additional  changes will be needed 
 in the parameter space to gracefully exit  the phase of pure de Sitter expansion, that will
  further modify  the scalar profile. We do not
consider this stage any further, since it occurs after the non-attractor regime we are interested in. 

  \smallskip
  
  So far we have assumed that our background corresponds to   a pure de Sitter universe, with constant Hubble parameter that we dub $H_0$ in all phases of evolution.
   Examining eq. \eqref{eomH}, we learn that for $t<t_0$, when the scalar time derivative is constant, the Friedmann equation can be satisfied
   with constant Hubble parameter $H_0$, by choosing the constant potential $V_1$ as follows 
   \be
 V_1\,=\,\frac{H_0^4}{18} \,\frac{\alpha_1^2 \left( -\rho_1+2\delta_1+3 \beta_1+4 \sigma_1\right)}{\left( -\rho_1+\beta_1+\delta_1+\sigma_1\right)^2}\,.
 \ee

 When $t>t_0$, the scalar field is rapidly varying,
  and at first sight 
  it seems difficult to satisfy the Friedmann eq.~\eqref{eomH} with $H(t)=H_0$ constant. But we can explore the non-linear structure
  of the kinetic sector of Horndeski action: we set the potential to zero,
 $V_2=0$, so that   the background evolution eq.~\eqref{eomH} becomes
 \be \label{eomH2}
0\,=\,\frac{\dot \phi^2}{2} \left( -\rho_2+2\,\delta_2\,\frac{H(t)}{H_0}+3\beta_2 \frac{H^2(t)}{H_0^2}
+4\sigma_2 \frac{H^3(t)}{H_0^3}
\right) \,.
\ee
 The parenthesis in the previous equation is a polynomial in the Hubble parameter $H(t)$, and we can select our parameters so that it admits a real constant
 root: the choice we  adopt  for this purpose is  
 \be\label{Hsol}
 \rho_2\,=
 2 \delta_2  +3\beta_2 + 4 \sigma_2 \,,
\ee
which makes the parenthesis in eq.~\eqref{eomH2} vanish when $H(t)=H_0$. This is a `self-accelerating' de Sitter branch of  solutions, where the non-standard kinetic terms allow us to have a pure de Sitter expansion also in a phase where the scalar field is rapidly varying.
%
  %

After determining the homogeneous background configurations for our system,
we now analyze the behaviour of tensor and scalar fluctuations.

\subsection{Dynamics of  fluctuations}

The dynamics of   primordial tensor and scalar fluctuations for inflationary models
based on Horndeski theory have  been explored in detail \cite{Kobayashi:2011nu}. See instead e.g. \cite{Adams:2001vc,Gong:2005jr} for systematic studies on the dynamics
of cosmological fluctuations in inflationary models with parameter discontinuities. 
Tensor fluctuations
  around a FRW background  metric with homogeneous scalar profile are controlled  by the following quadratic action
  
\bea\label{gt}
S^{(2)}_T
&=& \fr{1}{8}\int d t\,d^3 x\, a^3\,\left[  {\cal G}_T \left( \partial_t { h}_{ij} \right)^2-\frac{ {\cal F}_T}{a^2} \,\left( \vec \nabla h_{ij} \right)^2\right]\,,
\eea
where (recall we are focussing on kinetically driven scenarios, where the functions $G_a$ only depend on $X$)
\bea\label{tf}
{\cal G}_T&=&2\,\left[G_4-2 X G_{4X}-X 
H \dot{\phi} G_{5X}
\right]\,,
\\
{\cal F}_T&=&2\,\left[G_4-
 X 
\ddot{\phi} G_{5X}
\right].
\eea
Scalar fluctuations are described by the quadratic action 

\bea\label{gs}
S^{(2)}_S
&=& \fr{1}{2}\int d t\,d^3 x\, a^3\,\left[  {\cal G}_S \left( \partial_t { \cal R} \right)^2-\frac{ {\cal F}_S}{a^2} \,\left( \vec \nabla
\cal R \right)^2\right]\,,
\eea
where:
\bea\label{sf}
{\cal G}_S& = &\frac{\Sigma}{\Theta^2}\mathcal{G}_T^2 + 3\mathcal{G}_T \,,
\\
{\cal F}_S&=&\frac{1}{a}\frac{d}{dt}\left(\frac{a}{\Theta}\mathcal{G}_T^2\right)-\mathcal{F}_T\,,
\eea
and the explicit expressions of $\Sigma$ and $\Theta$ in terms of the functions $G_a$ are given in the  appendix \ref{A1}.
Tensor and scalar fluctuations are then
 characterised by the functions ${\cal G}_T$, ${\cal F}_T$, ${\cal G}_S$, ${\cal F}_S$ defined above. In the limit of very small $\alpha_2$
 we are adopting
\be
\alpha_2\ll1\,,
\ee
which is the relevant regime to have a phase of non-attractor inflationary evolution, 
 we find that each of these functions is proportional
to $\dot \phi^2/H_0^2$ in both phases of evolution,
 namely:
\bea
{\cal G}_{T i}&=&g_{t i}\,\frac{\dot \phi^2}{H_0^2}
\hskip0.8cm,\hskip0.8cm {\cal G}_{S i}\,=\, g_{s i}\,\frac{\dot \phi^2}{H_0^2}
\,,
\label{funcsG}
\\
{\cal F}_{T i}&=&f_{t i}\,\frac{\dot \phi^2}{H_0^2}
\hskip0.8cm,\hskip0.8cm
{\cal F}_{S i}\,=\,f_{s i}\,\frac{\dot \phi^2}{H_0^2}
\label{funcsF}
\,.
\eea
 The parameters $g_{ti}$, $g_{si}$,
$f_{ti}$, $f_{si}$,
  are constant, and we need
them to be positive to avoid instabilities. 

\smallskip

In the first part of the evolution, $t\le t_0$, we set $\delta_1=0$ (for simplicity) and  find  the following
expressions for the constant parameters entering eqs.~\eqref{funcsG}, \eqref{funcsF},
\bea
 f_{t1}&=&-\frac{\beta_1}{ 6}\,,
\\
g_{t1}&=&\frac{\beta_1+3\sigma_1}{6}\,,
\\
\label{gs1}g_{s1}&=&\frac{\left(\beta_1+3\sigma_1\right) \left( 
3 \beta_1^2+\rho_1(\beta_1 + 3\sigma_1)+2\sigma_1(4\beta_1 + 3\sigma_1)
\right)}{18 (\beta_1+2 \sigma_1)^2}\,,
\eea
while the quantity $f_{s1}$ satisfies 
the condition
\be
\label{fs1}f_{s1}\,=\,\frac{\left( f_{t1}-g_{t1}\right)^2}{2 g_{t1}-f_{t1}}\,,
\ee
which relates its value to the remaining quantities. {The stability of fluctuations in the first phase of the evolution thus  requires the following relations: $\beta_1 < 0$, $\beta_1 + 3\sigma_1 >0 $ (implying $\sigma_1 >0$) and $\beta_1 + 3\sigma_1 > -\beta_1/2$. With these choices, one can easily satisfy $g_{s1}>0$ by choosing $\rho_1$ appropriately.} 

\smallskip

The second part of the evolution, $t_0\,<\,t \,<\, t_1$, is the most interesting {since $\dot \phi\propto a^{-3}$} and we are in a non-attractor phase
 which enhances tensor fluctuations.
  In this case, we find that the parameters of eqs.~\eqref{funcsG}, \eqref{funcsF} read
\bea
 f_{t2}&=&-\frac{\beta_2+9 \sigma_2}{6}
 \,,
 \\
 g_{t2}&=&\frac{\beta_2 + 3\sigma_2}{6}
\,,
\\
\label{fs2}f_{s2}&=&\frac{ -2 \beta_2^2 +\delta_2 (\beta_2 + 9\sigma_2) + 3\sigma_2 (\beta_2 + 3\sigma_2)
} {6 \left( 3\beta_2+\delta_2+6 \sigma_2\right)}
\,.
\eea
Moreover,
 we find the relations
\bea
\label{gs2}g_{s2}&=&\frac{3}{5}\,\left(f_{s2}+f_{t2}+5 g_{t2}\right)
\,,
\\
\rho_2&=&-\frac{f_{t2}
\left(f_{s2}+f_{t2}\right)
+19 g_{t2} \left(f_{s2}+f_{t2}
 \right)
+60 g_{t2}^2
}{f_{s2}+f_{t2}}\,,
\eea
where we  used \eqref{Hsol}. Note that these relations imply $\rho_2 <0$,  and hence $g_{s2} >0$ whenever $f_{t2},g_{t2},f_{s2} >0$. Imposing
 the condition  \eqref{Hsol} in the second phase of the evolution, we  require the following relations to hold in order to avoid instabilities of the tensor and scalar fluctuations: $\beta_2 >0$, $\sigma_2 < 0$, $\delta_2 < 0$ where $\beta_2 + 9\sigma_2 < 0$, $\beta_2 + 3\sigma_2 > 0$ and $f_{s2} > 0$ by choosing an appropriate $\delta_2$.

 
 As we explained above, we are not interested in investigating the last stage of slow-roll evolution, for $t\ge  t_1$ (see eq.~\eqref{defttil}) when the size of the parameter $\alpha_2$
 becomes important. This phase occurs {\it after} the non-attractor phase we wish to investigate, and the properties of the system
and the  evolution of fluctuations can change and are not  necessarily described by the action we are interested in. 

We
now analyse  an explicit choice of quantities
within the available  parameter space which satisfies the aforementioned  stability conditions, and discuss the corresponding 
physical consequences. See also appendix \ref{sec-disf} for an investigation of our system in a `Einstein frame' where tensor fluctuation at quadratic order obey a standard action.

\begin{figure}[h!]
\begin{center}
\includegraphics[width = 0.49 \textwidth]{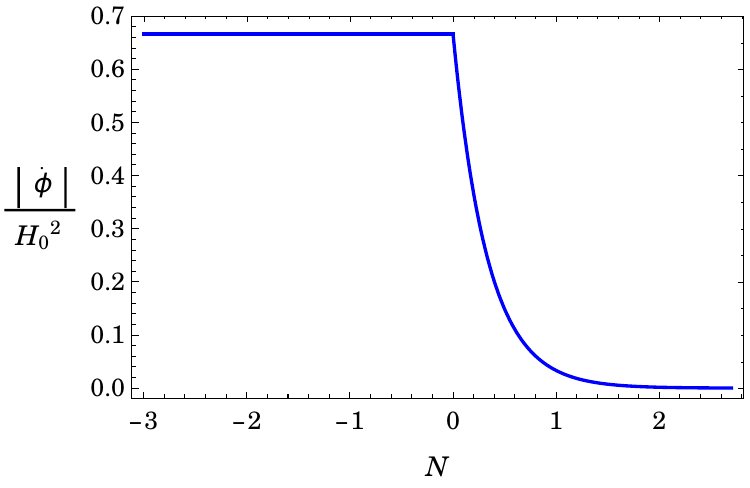}
\includegraphics[width = 0.49 \textwidth]{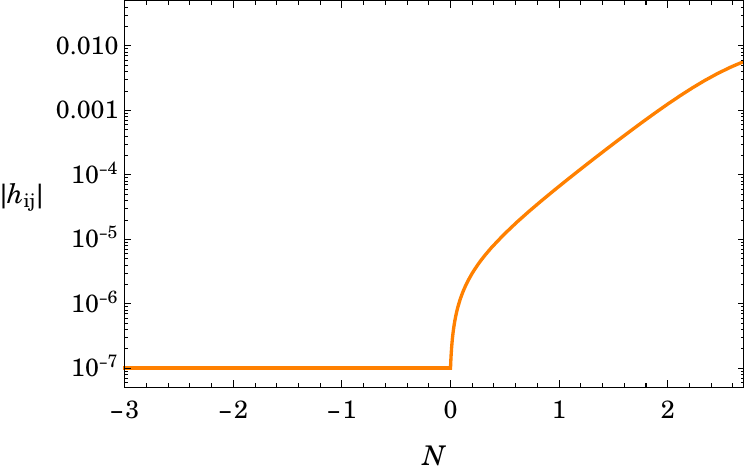}
 \caption{\it 
The behavior of $\dot \phi$ and $|h_{ij}|$ through the transition to the non-attractor phase where we used the expressions \eqref{dphi} and \eqref{solsh1} respectively. On the right, we have set $\mathcal{C}_2 = \mathcal{C}_1 (H_0/c_T)$ with $\mathcal{C}_1 = 10^{-7}$ in \eqref{solsh1} to show the evolution of $|h_{ij}|$. For both plots, we have the following choice of parameters:  $\alpha_1=1,\,\beta_1=-1,\,\sigma_1=2.5,\,\rho_1=1,\,\delta_1\,=\,0$, $\alpha_2\,=\,10^{-3},\,\beta_2\,=\,1.5,\,\sigma_2\,=\,-0.2,\,\delta_2\,=\,-4,\,\rho_2\,=\,-4.3$ where $N=0$ corresponds
to the transition point. In these plots, the non-attractor phase lasts approximately $\Delta N \simeq 2.7$ e-folds. }
\label{fig:scal}
\end{center} 
\end{figure}

\subsubsection*{Enhancement of  spectra of fluctuations during the non-attractor phase} 
 
 In the non-attractor phase we are considering, both tensor and scalar fluctuations
can be enhanced.  
The   
  tensor  pump field during the non-attractor phase is given by
 \be
\tilde{z}_T = \fr{a}{2}\left(
{\cal G}_T {\cal F}_T
\right)^{1/4} \,=\,\frac{a \dot \phi}{2H_0}\,(g_{t2} f_{t2})^{1/4}\,,
\ee
whereas the scalar pump field is given by the following expression:
\be
\tilde{z}_S = \sqrt{2} a \left(
{\cal G}_S {\cal F}_S
\right)^{1/4} = \,\frac{\sqrt{2} a\dot \phi}{H_0}\,(g_{s2} f_{s2})^{1/4}\,.
\ee
In both of the expressions for $\tilde{z}_T$ and $\tilde{z}_S$, the scalar field velocity reads\footnote{Recall that we are assuming $\alpha_2 \ll 1$.} 
\beq
\dot{\phi} \approx -\frac{H_0^2\,\alpha_1}{3 \left(-\rho_1+\delta_1+\beta_1+\sigma_1 \right)} \,a^{-3}\,.
\eeq 
As
$\dot \phi \propto a^{-3}$, we realise the condition \eqref{con2o} we identified before. We therefore realize the tensor 
duality and as can be anticipated from the expression in \eqref{tensenha}, we will have an enhancement of the tensor power spectrum: modes
that leave the horizon within the time interval $t_0\,<\, t\,<\, t_1$ corresponding to the non-attractor phase will receive an exponential enhancement proportional to $a^{6}$.{ Notice that since fluctuations evolve beyond the Hubble horizon, the tensor and scalar power spectra should be evaluated at the end of the non-attractor phase which we denote by $t_1$. Therefore, using \eqref{defttil}, the tensor power spectrum at the end of the non-attractor phase can be amplified with respect to the one during the preceeding slow-roll by the following amount
\bea\label{TE}
\nn \frac{{\cal P}_{\tilde h}}{{\cal P}_{ h}}~~ \bigg\rvert_{t = t_1}\,=\,\left(\frac{z_T}{\tilde z_T}\right)^2\,\bigg\rvert_{t = t_1} &=& a(t_1)^{6} \left(\frac{g_{t1} f_{t1}}{g_{t2} f_{t2}}\right)^{1/2}\\
&  \simeq& 
\left(\frac{\alpha_1}{\alpha_2}\right)^2 \,\frac{( -\rho_2+\delta_2+\beta_2+\sigma_2)^2}{(-\rho_1+\delta_1+\beta_1 +\sigma_1)^2} \left(\fr{\beta_1 (\beta_1 + 3\sigma_1)}{(\beta_2+3\sigma_2)(\beta_2+9\sigma_2)}\right)^{1/2}.
\eea
It is worth emphasizing again that we can enhance the power in the fluctuations while maintaining a scale invariant statistics deep in the non-attractor regime\footnote{At scales corresponding to the modes leaving the horizon slightly before the transition to non-attractor regime, we expect peaks or features in the fluctuation power spectra, whose study require more careful numerical investigations, see \eg \cite{Leach:2001zf}.}. This behavior can be particularly interesting to build models with an enhanced
tensor power spectrum detectable at interferometer scales, see the brief  discussion in section \ref{sec-out}. See also \cite{Jain:2009pm} for a scenario
able to amplify the tensor-to-scalar ratio $r$  at small scales, by reducing the size of the spectrum of scalar fluctuations.
}

Similarly to the tensor fluctuations, scalar fluctuations grow during the non-attractor regime since the scalar pump field 
 $\tilde{z}_S$ has a similar
structure to $\tilde{z}_T$. Using Wands' duality (see footnote \ref{fot1}) we have
\beq\label{SE}
\frac{{\cal P}_{\tilde{\mathcal{R}}}}{{\cal P}_{ \mathcal{R}}}~~ \bigg\rvert_{t = t_1}\,=\,\left(\frac{z_S}{\tilde z_S}\right)^2\,\bigg\rvert_{t = t_1} = a^6(t_1)\, \left(\frac{g_{s1} f_{s1}}{g_{s2} f_{s2}}\right)^{1/2},
\eeq
where $g_{s1}$, $f_{s1}$, $f_{s2}$ and $g_{s2}$ defined as in  \eqref{gs1},\eqref{fs1}, \eqref{fs2} and \eqref{gs2} respectively.

As a specific example, we take the following set of parameters 
\be
\left(\label{exmp}
{\alpha_i,\beta_i,\,\delta_i,\,\rho_i,\,\sigma_i}
\right)
\,=\, \left\{ \begin{array}{l} 
\left({1,\,-1,\,0,\,1,\,2.5}\right) \hskip4.9cm t\,<\,t_0
\\ 
\\
 \left({10^{-3},\,1.5,\,-4,\,-4.3,\,-0.2}\right) \hskip3cm t\,>\,t_0.
 \end{array} \right. 
 \ee
{For this choice, the corresponding field velocity $\dot{\phi}$ and the amplification of $|h_{ij}|$ using the super-horizon expression in \eqref{solsh1} are shown in Figure \ref{fig:scal}. However, we reiterate that equation \eqref{solsh1} can be only used as a rough indicator of the enhancement of the fluctuations as it relies on strict super-horizon limit $k = 0$ (see \cite{Leach:2001zf}	). On the other hand, to obtain a more accurate estimate on the enhancement we use the expressions based on the duality in \eqref{TE} and \eqref{SE} with the parameter choices in \eqref{exmp}. We find that the tensor power spectrum can be enhanced by a factor of $\approx 5 \times 10^{7}$ and the scalar power spectrum by a factor of $\approx 5 \times 10^{6}$ while the system satisfies all the stability constraints in both sectors\footnote{Interestingly, we learn that scalar fluctuations are less enhanced than tensor ones. This could be useful when building more realistic scenarios of our mechanism,  to avoid constraints from excessive primordial black hole production.}. In general, looking at the structure of the equations \eqref{SE} and \eqref{TE}, we see that the level of enhancement of the scalar and tensor power spectrum is mainly controlled by the duration of the non-attractor regime, in particular by the ratio $\alpha_1/\alpha_2$ whereas the other parameters in the model mainly serve to satisfy the stability conditions of the fluctuations.} 

We finally comment on the propagation speed of the fluctuations in the model we consider in this paper. As both the scalar and tensor fluctuations have kinetic functions that satisfy the relations: ${ \cal F}_T \neq {\cal G}_T$ and ${\cal F}_S \neq {\cal G}_S$, they exhibit non-trivial sound speeds given by the following expressions 
\be
c_S^2
\,=\, \left\{ \begin{array}{l} 
f_{s1}/g_{s1}\,, \hskip2cm t\,<\,t_0\,,
\\ 
\\
f_{s2}/g_{s2}\,, \hskip2cm t\,>\,t_0\,,
 \end{array} \right. 
 \ee
and 
\be
c_T^2
\,=\, \left\{ \begin{array}{l} 
f_{t1}/g_{t1} \,,\hskip2cm t\,<\,t_0\,,
\\ 
\\
f_{t2}/g_{t2}\,, \hskip2cm t\,>\,t_0\,.
 \end{array} \right. 
 \ee
For the specific parameter choices we made in \eqref{exmp}, we learn that during both phases of the inflationary
 evolution, for $t<t_0$ and $t>t_0$,  the scalar and tensor sound speed is less than unity, i.e $c_T^2 < 1$ and $c_S^2 < 1$. In particular, in the slow-roll phase, the tensor sound speed is $c_T^2 \approx 0.15$ and it increases to $c_T^2 = 0.33$ during the non-attractor phase. Similarly, scalar sound speed is given by $c_S^2 \approx 0.69$ during the slow-roll era and increases to $c_S^2 \approx 0.89$ in the non-attractor era.

\smallskip
\subsubsection*{Non-gaussianity}
We conclude this section with some comments on tensor non-Gaussianity,
an observable that can be useful for discriminating among primordial and astrophysical stochastic gravitational wave background detectable with interferometers
\cite{Bartolo:2018qqn}.
Tensor non-Gaussianity  is also an important observable for charactizing the primordial 
stochastic gravitational wave background at CMB scales, and have been explored in other contexts, see e.g.  \cite{Thorne:2017jft,Agrawal:2017awz,Agrawal:2018mrg}. 
We briefly consider   tensor non-Gaussianity with a shape enhanced in the squeezed limit: other shapes of tensor non-Gaussianity can be produced in single field
inflation -- see \cite{Gao:2011vs} -- but we do not consider them in this context.
This shape is controlled
by the following third order action, obtained by expanding up to third order in tensor fluctuations
the Horndeski set-up we are examining
 \cite{Gao:2011vs}
\be
S_{T}^{(3)}\,=\,\int dt \,d^3 x\,a\,
\frac{{\cal F}_T}{4 }\,\left( 
h_{ik } h_{jl}-\frac12 h_{ij} h_{kl}\right)\,\partial_k \partial_l\,h_{ij}
\,.
\ee
Considering the tensor dual of a slow-roll phase as described in section \ref{sec-dua}, we learn that in terms of the tensor dual variable
$ \tilde h_{ij}\,=\,\left( z_T/\tilde z_T\right)\,h_{ij}$ the overall factor in the previous third order action changes to  
\be
S_{\tilde T}^{(3)}\,=\,\int dt \,d^3 x\,a\,
\frac{{\cal F}_T}{4 }\,\left(\frac{ z_T}{\tilde z_T}\right)^3\,\left( 
\tilde h_{ik } \tilde h_{jl}-\frac12 \tilde h_{ij} \tilde h_{kl}\right)\,\partial_k \partial_l\,\tilde h_{ij}\,.
\ee
Hence, if the ratio $ z_T/\tilde z_T$ is large as for the tensor dual of a slow-roll phase, see Section \ref{dualsr},  the amplitude of 
squeezed tensor non-Gaussianity can increase in the tensor dual regime\footnote{See also \cite{Ricciardone:2017kre} for other models with large squeezed tensor non-Gaussianity.}. It would be interesting to develop further this subject, and investigate its consequences
for the detectability of tensor non-Gaussianity at interferometers, as explored in \cite{Bartolo:2018qqn}. We plan to do so in a forthcoming work. 

\section{Outlook}\label{sec-out}

In this work we discussed a new mechanism to amplify tensor fluctuations during single field inflation, by exploiting
a phase of non-attractor evolution. {We have identified the necessary condition for amplifying  the tensor spectrum  at super-horizon scales, which  is that the tensor pump field, defined in eq. \eqref{comz}, decreases with time during a phase of the inflationary
evolution. The would-be decaying tensor mode gets then enhanced and increases the size of tensor fluctuations.  
 We determined a criterium, which we dub tensor duality, that allows us to 
analytically estimate the statistical properties of the amplified tensor fluctuations during the non-attractor
era. We then built and investigated in detail a concrete model of kinetically driven inflation able to
satisfy our conditions, and analytically determined the properties of the enhanced spectrum of tensor modes in this set-up.
Much work is left for the future:
\begin{itemize}
\item Our concrete scenario is based on G-inflation, since we need a non-trivial kinetic mixing between scalar
and metric to realise our mechanism. It will be interesting to understand whether other realisations can exist,
for example by means of sudden changes in the tensor sound speed due to effects of new heavy physics or string
theory, as 
in \cite{Achucarro:2010da}. 
\item  The quadratic tensor action we obtained in our system is distinct from the one of single field inflation
with standard kinetic terms, but a sequence of conformal and disformal transformations can recast
it in standard form \cite{Creminelli:2014wna}. In Appendix \ref{sec-disf} we show that our system, during the non-attractor
era we have investigated, can be disformally related to a rapidly contracting universe.  It will be interesting
to further explore the physical implications of disformal transformations during non-attractor regimes.
\item We provided evidence that the spectrum of tensor fluctuations can be non-Gaussian, besides being enhanced. It will be important to analytically study in more details the amplitude and shape of non-Gaussianity of tensor
modes in our set-up. 
 \item Finally, it will be important to build a complete, realistic  scenario (based on G-inflation or on other theories) able to 
 sufficiently amplify tensor modes at interferometer scales, and study prospects for the detectability of the stochastic
 primordial tensor background and its non-Gaussianity.
 \end{itemize}
We hope to further report soon on  these topics.

\subsection*{Acknowledgments}

MM, O\"O, GT and IZ are partially supported by the STFC grant ST/P00055X/1. 

\begin{appendix}

\section{ Background equations in Generalized G-Inflation and functions $\Sigma$ and $\Theta$}\label{A1}
In a FRW universe, the equations describing the background evolution can be written in an analogous way to the minimally coupled canonical scalar field. Since we are dealing with a shift symmetric system, the background equations are simpler compared to the general expressions given in \cite{Kobayashi:2011nu}. In this case, the generalized energy and the sum of energy and pressure densities can be expressed in the following way 
\begin{align}\label{con1}
\dot{\phi}J-G_{2}+ 2X\left[6H^2G_{4X}+ 2H^3\dot{\phi}G_{5X}\right]	=6H^2 G_4,
\end{align}
\beq
\label{con2}\dot{\phi}J -2X\ddot{\phi}G_{3X}+ 2\fr{\d}{\d t}\left[2H(G_4-2XG_{4X})-H^2X\dot{\phi}G_{5X}\right]=0,
\eeq
where
\beq\label{J}
J = \dot{\phi} \left[G_{2X} + 3H\dot{\phi}G_{3X}+6H^2(G_{4X}+2XG_{4XX})+H^3\dot{\phi}(3G_{5X}+2XG_{5XX})\right]
\eeq
The generalized Friedman equations above can be combined with the scalar-field equation 
\beq\label{Jeq}
\fr{1}{a^3}\fr{\d}{\d t}(a^3 J) = 0\,,
\eeq
to close the system of equations. On the other hand, the functions $\Sigma$ and $\Theta$ appearing in the kinetic functions $\mathcal{F}_S$ and $\mathcal{G}_S$ \eqref{sf} of the scalar fluctuations are given by the following expressions
\begin{align}
\nn \Sigma \equiv X G_{2X} + 2X^2 G_{2XX}&+ 12H\dot{\phi}XG_{3X}+6H\dot{\phi}X^2 G_{3XX} - 6H^2G_{4}\\
\nn &+6H^2 (7XG_{4X} + 16X^2 G_{4XX} + 4X^3 G_{4XXX})\\
&+ 30H^3\dot{\phi}XG_{5X} + 26H^3\dot{\phi}X^2 G_{5XX}+ 4H^3\dot{\phi}X^3G_{5XXX},
\end{align}
and
\begin{align}
\Theta = -\dot{\phi}XG_{3X} + 2HG_4-8HXG_{4X}-8HX^2G_{4XX}-H^2\dot{\phi}(5XG_{5X}+2X^2G_{5XX}).
\end{align}

\section{ Disformal and conformal transformation of the tensor action}
\label{sec-disf}
 
 In  \cite{Creminelli:2014wna} it has been shown that a combination of  conformal and disformal transformations
 allows one to recast the quadratic tensorial action into a form identical to the action of
  tensor modes in  standard slow-roll
 inflation. See also \cite{Domenech:2015hka} for an analysis of the consequences of disformal transformation
 for cosmological fluctuations. We discuss the implications of such transformations for our set-up.

We follow the prescription of \cite{Creminelli:2014wna,Baumann:2015xxa} in defining the conformal and disformal transformations, and we
redefine the time coordinate and scale factor as

\begin{equation}\begin{split}
\dd {t} = (c_T \mathcal{F}_T)^{-1/2}~ \dd{\hat{t}}, \quad  a= \left(\frac{c_T}{ \mathcal{F}_T}\right)^{1/2} \hat{a}.
\label{eq:102}
\end{split}\end{equation} 
(here $c_T^2\,=\,\mathcal{F}_T/\mathcal{G}_T$) to ensure the metric acquires a standard FRW form.
With these redefinitions, the tensor action now reads 

\begin{equation}\begin{split}
S&=\frac{1}{8} \int d{\hat{t}}\, d^3{x} \, \hat{a}^3  \bigg[(\partial_{\hat{t}} h_{ij})^2 - (\nabla h_{ij})^2\bigg].
\label{eq:103}
\end{split}\end{equation}
as in standard single field inflation in an Einstein frame. Using relations \eqref{eq:102}, we can compare the
 Hubble parameter associated with the new scale factor $\hat a (\hat t)$ 
with the quantities defined in terms of the original time $t$:
\be
 \hat{H} = \frac{1}{\hat a}\frac{d\, \hat{a}}{d \hat{t}} 
 =  (c_T\mathcal{F}_T)^{-1/2} \left[ H + \frac{1}{2} 
 \left( \frac{ \dot{\mathcal{F}_T}}{\mathcal{F}_T}-\frac{\dot c_T}{c_T}  
 \right)  
 \right],
\label{eq:105}
\ee
We then use the structure of the relations \eqref{funcsG}, \eqref{funcsF} to evaluate the right hand side of \eqref{eq:105}. 
When focussing on the first phase of slow-roll regime, we find, as expected, that the new Hubble
parameter $\hat H$ is proportional to the original one $H$. 

But when evaluated in the non-attractor phase, using the relations (see eqs \eqref{funcsG} and \eqref{funcsF})
\beq
{\cal F}_{T}=f_{t 2}\,\frac{\dot \phi^2}{H_0^2},~~~~~~{\cal G}_{T}=g_{t 2}\,\frac{\dot \phi^2}{H_0^2},~~~~~~c_T=\sqrt{\frac{f_{t2}}{g_{t2}}}
\eeq
and the fact that $\dot \phi \,=\, \phi_0/a^3$ for a constant $\phi_0$, we find the following expresion for the Hubble parameter in the Einstein frame
 \bea
 \hat H\,&=&\,-\frac{2\,g_{t2}^{1/4}H^2\,a^3}{\,f_{t2}^{3/4}\,|\phi_0|}\,,\\
 &=& - \sqrt{\frac{4H\phi_0 g_{t2}^{5/4}}{f_{t2}^{3/4}}}~~\hat{a}^{-3/2},
 \eea
 where in the second line we used the second relation in \eqref{eq:102} between the scale factors in the two frames. Hence in the Einstein frame where tensor fluctuations are controlled by action \eqref{eq:103}, the background geometry in the non-attractor regime is
 described in terms of a dust dominated {\it contracting} universe.
 
 This impliest that, within the Einstein frame description developed in this appendix, the Universe undergoes a short phase of contraction - lasting a few e-folds - during which the amplitude of tensor fluctuations can grow. This perspective offers another point of view for the results in the main text, within a frame where the action for quadratic tensor fluctuations is standard. It would be interesting to embed our scenario in a set-up with smooth transition between expanding and contracting phases, and study in detail the matching and stability issues for fluctuations. Possible instabilities in the bouncing transition phase can be tamed in sufficiently rich scalar-tensor systems related to the set-up we use in this work. A detailed analysis of this subject is beyond the scope of this article, and we leave it for  future investigations.
\end{appendix}

\bibliographystyle{utphys}

\bibliography{refs}

\end{document}